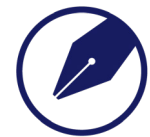

Research Article

# Fostering Student Engagement in a Mobile Formative Assessment System for High-School Economics

Fotis Lazarinis[1*] , Dimitris Kanellopoulos[2]

[1]Hellenic Open University, GR 26335, Patras, Greece
[2]Department of Mathematics, University of Patras, GR 26500, Patras, Greece
 Email: fotis.lazarinis@ac.eap.gr

**Abstract:** In a mobile learning environment, students can learn via mobile devices without being limited by time and space. Therefore, it is vital to develop tools to assist students to learn and assess their knowledge in such environments. This paper presents a tool/application for formative self-assessment. The tool supports the selection of questions based on user-defined criteria concerning (1) the difficulty level; (2) the associated concepts; and (3) the purposes of the test taker. The main purpose of the presented tool is to better support the learning aims of the participants and to increase their engagement in the learning process. The focus of this study is to evaluate the tool using quizzes in Microeconomics to realize its potential in this specific domain. Teachers and students were involved in the experiments conducted. The experiments demonstrated that the presented tool is usable; it motivates the students and improves their understanding.

## 1. Introduction

In the 21st century, educational tools are utilized in various domains as a prime source of teaching or supplementary to the main teaching process. In particular, information and communication technologies (ICTs) are employed in various educational actions. The benefits of ICTs adoption in education and of e-learning are well-documented in many studies (e.g., Basak et al., 2018; Welsh et al., 2003). ICTs have been integrated into many educational activities with certain advantages (Livingstone, 2012). These advantages range from access from anywhere/anytime to enhanced multimedia representations to the adaptation of the teaching materials and the process to the needs of the individual (i.e., personalization of content delivery) (Christudas et al., 2018).

In such e-learning contexts, assessment plays an important role in every learning activity (Ross & Morrison, 2009). Formative assessment has been defined as "the process used by teachers and students to recognise and respond to the students' learning in order to enhance that learning, during the learning" (Cowie & Bell, 1999). Under this broad definition, different forms of assessment emerged as such as self, peer, collaborative, and goal-based assessment. Formative assessment is important for self-regulated learning (Clark, 2012). Self-assessment is an important type of formative assessment for two main reasons. First, because it happens before every learning process. Students, more often than not, self-assess their knowledge before attending a class or before a final assessment. Secondly, self-assessment helps students to evaluate their level of understanding and eventually to promote the learning process (Andrade & Valtcheva, 2009). This kind of assessment is often informal and ad hoc. However, it is an integral part of the learning process, and therefore it has to be treated more systematically. The use of self-assessment increases students' metacognition positively and impacts students' learning and self-regulation (Siegesmund, 2017). This happens because the ability of students to effectively self-regulate their learning is dependent on their metacognitive ability. Undoubtedly, self-assessment is an integral part of the learning process, and therefore it has to be treated more systematically.

Mobile learning (m-learning) is considered as a new stage of e-learning (Georgiev et al., 2004). In the last two decades, numerous higher education institutions support their traditional learning modes with various mobile learning initiatives. It can be said that m-learning is becoming gradually more universal (Mehdipour & Zerehkafi, 2013). This fact emphasizes the potential need for mobile learning policies that support higher education institutions to navigate growing globalization. As a result, a lot of mobile learning tools have been proposed for various fields such as grammar learning (Shuib et al., 2015), anatomy teaching (Golenhofen et al., 2020), and Microeconomics (Muslimin et al., 2017).

Education in Economics is an area that could greatly benefit from the adoption of ICT technologies, given the pace that multimedia data on economics (e.g., video presentations) are being continuously produced (Kuhn et al., 2018). E-learning has been applied in the teaching of Microeconomics (e.g., Kamarni & Rahadian, 2021; Jiang, 2020; Kuhn et al., 2018; Chen & Lin, 2015; Novo-Corti et al., 2013) with positive outcomes. For instance, Muslimin et al. (2017) designed and developed a mobile educational application (called MobiEko Apps) for learning Microeconomics concepts which constitute part of an education curriculum. Moreover, they evaluated their tool regarding its learning effectiveness and acceptance in relation to the presentation, visual, navigation, and accessibility of mobile app design. Meanwhile, usability assessment models for mobile learning tools are conducted from multiple approaches, mainly the technological and pedagogical, and from different actors: students, teachers, and experts. However, there is limited research work focused on the usability assessment of mobile learning tools through field studies with students and teachers in a real environment (Chiu et al., 2018; Huang & Chiu, 2015).

In this paper, we present a mobile learning tool that allows high school students to self-assess their knowledge on Microeconomics. The field of our study is Microeconomics. We consider the appropriateness of our application for learning Microeconomics concepts. In particular, we focus on introductory Microeconomics to help students to comprehend certain Microeconomics concepts related mainly to demand and price elasticity. To achieve this, we resort to formative assessments in a mobile environment. Moreover, we present a usability assessment of the mobile learning tool based on Microeconomics with ten teachers and eighty-three high-school students in a real environment. The results obtained from the usability assessment show that the presented tool significantly improves the motivation and participation of students in the learning process.

## 2. Literature review

ICT technologies have been used in Microeconomics instruction from early on. For example, Boyd (1998) used the graphical capabilities of a symbolic processing program to help students understand microeconomic functional relationships which would be more difficult through a textbook. The use of spreadsheets in the Microeconomics curriculum has been explored in (Clark & Hegji, 1997). The benefits of using Internet resources to enhance economic courses have been identified from the early days of the Internet (Agarwal & Day, 1998).

From these early days onwards, several manifestations of technology integration into the economics curriculum have appeared. In Novo-Corti et al. (2013) a methodology combining the assessment with multiple-choice tests through the virtual environment Moodle and the evaluation by using the traditional classroom exams is discussed. Their research showed that the frequent interaction of the students with the e-learning tool, helped students to be more prepared for the final exam. Further, students were quite satisfied with the e-learning approach and agreed on the effectiveness of ICTs as a means of learning in the field of Microeconomics. In a similar study, concerning public administration topics, the authors measured a statistically significant increase of students' effectiveness, calculated with the average grade and the average number of admissions to the exams (Umek et al., 2015).

The online behavior of college students in an online course has been studied in (Chen & Lin, 2015). The authors analyzed data from an online undergraduate intermediate Microeconomics course. They recorded different learning patterns of the students. The researchers discovered that even the students who studied the online materials only before the exams had comparable performance to those who used a more balanced study approach through frequent visits to the online course. In general, students who finally used the e-learning tool were benefited.

A recent study states that "technology provides the vehicle for economics education to break free of the constraints of monist teaching methods and ensures that economics students can fully engage in the discipline's vibrant debates" (Watson & Parker, 2016). The researchers argue that a blended learning approach should play a central role in economics education. Flipped learning has also been proposed as a method to increase students' understanding and participation in economics (Roach, 2014). Improved students' learning has been demonstrated when courses were taught in the flipped format, i.e. with the aid of educational technologies. Svoboda et al. (2016) recorded an improvement in the performance of students enrolled in an e-course of economics realized through Moodle.

E-assessment plays an important part in learning as it can broaden the range of what it can be tested. It has been shown that feedback from e-assessment adds value to learning (Dermo, 2009). Self-assessment techniques have been applied on a large scale in massive open online courses (MOOCs) and have been proved beneficial (Kulkarni et al., 2013).

Formative assessments as an instructional tool in economics have been proposed in (Walstad et al., 2010). Student-generated material has been used as supplemental teaching material. That way more focused feedback and instructions can be given to students. Conventional take-home formative tests and online quizzes are compared using data from six semesters of an economics course (Maclean & McKeown, 2013). Both approaches aimed to engage students in learning and the characteristics of each approach were considered. The main finding is that formative assessments implemented as online quizzes are as effective as take-home assignments but their cost is considerably less. *'Challenge quiz'* is a tool to support learning by students. The tool has been used in two US institutions in courses on the principles of Microeconomics. The analysis of the data indicates that students taking those quizzes tend to engage in more study methods and perform better than their initial in-class quiz, despite the increased rigor of the assessment (McGoldrick & Schuhmann, 2013).

Mobile versions of self-assessments have been already implemented. Bogdanović et al. (2014) showed that the integration of the mobile quiz application into Moodle improves students' results and increases satisfaction and motivation for using mobile devices in their learning process. Hwang et al. (2011) proposed a formative assessment-based approach for improving the learning achievements of students in a mobile learning environment. Their experiments showed that their proposed approach promoted students' learning interest and attitude and improved their learning achievement.

The above studies show that economics education can benefit from educational technologies in general and self-assessment in particular. Mobile applications for effective support of self-assessments could reduce the cost of the assessment, enrich the process with multimedia representations, provide instant and personalized feedback, enhance the access mode, and help more students to be actively engaged.

## 3. Self-assessments through a mobile application

In this study, we focus on the usage of a mobile application for assessing concepts of Microeconomics in highschool education. The design principles and the first version of the utilized mobile assessment tool are discussed in (Lazarinis et al., 2017). In the current paper, we briefly present the new options of the tool and focus on its usage for self-assessments of Microeconomics concepts.

The self-assessment tool offers some adaptive options to the learners in order to customize the learning process to their interests and needs. Learners adapt their process to their current learning goals, their knowledge level, and the difficulty level of the assessment items. The tool is domain-independent and although it has been mainly used to assessing computer science concepts, in this work it has been used in economics education. Our main goals were to test its suitability for other domains and to measure the learning advantages of the students who used it.

Students can use the system either as registered users or as guests. Registration offers some additional options, such as assessment history, student profiles where the knowledge on various topics is kept, etc.

Figure 1 shows the basic options of the tool. Learners select one or more topics; define their knowledge level and the level of completed education on the selected topics; define the characteristics of the assessment items they want to try; define the number of items and finally execute the assessment. Based on the knowledge level and the learner's performance, some inferences about the knowledge of the test participants in the specific topics are possible.

The current version of the system supports multiple types of questions:
Multiple Choice
Multiple Response
True or False
Fill in the Blanks
Matching
Sequence
Hotspot
Drag and Drop
Select from Lists
Likert Scale

The question itself can be in text form, image, video or animation. These ways students have visualized information in assessment items which is very useful for many domains. Given that the tool is mainly for self-assessment, rich interactive experiences help the students to comprehend the presented concepts. Figure 2 shows an example of questions for Microeconomics. Each question is associated with a specific concept from a scientific field.

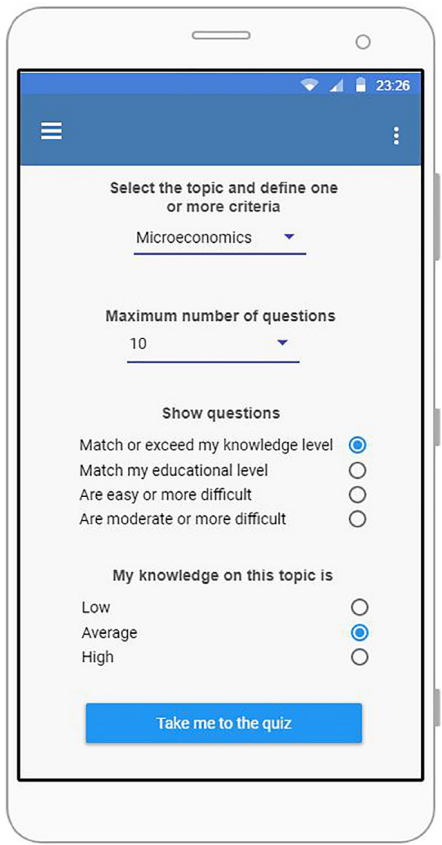
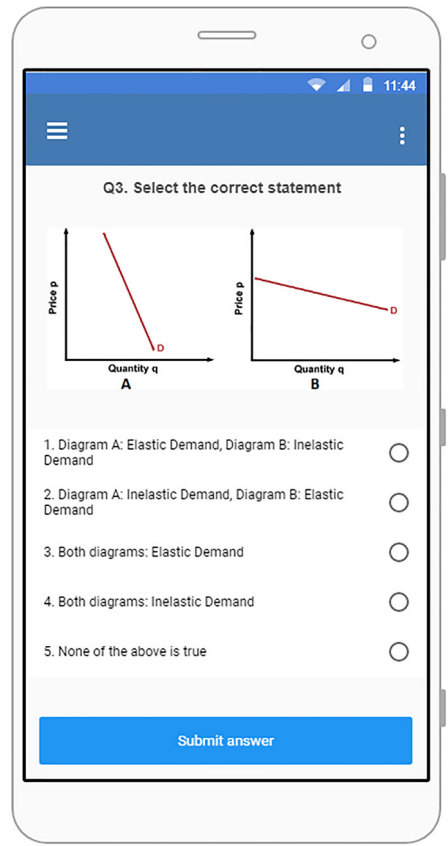

Figure 1. Self-assessment criteria  Figure 2. Sample question

At first, administrators have to define or update a topic hierarchy. As a result, for every new domain, a hierarchy is built and is attached to the main hierarchy. Then, educators enrich the item bank with questions, associate them with one or more topics and define the difficulty and related education level. Administrators and educators log in through a Web interface and update the question database. They define the text of the question, the possible answers, the correct one(s), the difficulty level, and the educational level. They match each question to a broad topic or a narrow subtopic.

In any case, the question is automatically assigned to the topics at the higher level of the hierarchy. This supports more clustering alternatives which could eventually improve the self-testing process.

Students fulfill their assessment goals by selecting the most appropriate items to be tested on and defining one of the followings:

1. The level of difficulty and/or the appropriate educational level of the testing items: Questions are classified as easy, medium, or difficult, while learners can select assessment items based on their difficulty. In particular, they can select questions that match or pass or are below a specific level of difficulty, e.g., "show questions that are above average difficulty".

2. The learners' knowledge level: students can select questions that match or are above or below their knowledge level. Easy, medium, and difficult questions correspond to low, good, and high knowledge levels. For example, students who have a "good" knowledge of a topic can define rules like "show questions greater or equal to my knowledge level". In that case, the tool retrieves testing items with a high difficulty.

3. The learners' educational level: this possibility supports the selection of questions based on the educational level they correspond to. Users can instruct the system to "retrieve the testing items that match their level of education".

Alternatively, students may ask the tool to decide on the arrangement of testing items based on the user-provided data of registered users. The application retrieves items that correspond to the students' educational level on a specific topic. The questions are grouped by subtopic and their level of complexity. Next, the questions matching the user criteria are presented to the student. The retrieved testing items are clustered by the subtopic they relate to.

Once a student completes the self-evaluation of her/his knowledge, the tool estimates the knowledge level and presents statistics per topic. For registered users, these statistics are stored in the database to be utilized in future assessments for this user. The new level of knowledge is based on the average score on each topic. Scores are normalized based on their question weights. A mark below 50% results in "low" understanding while good knowledge means a score between 50% and 75%. A result greater than these limits leads to a high knowledge level. Further, statistics per topic and subtopic are presented to students who are informed of their weaknesses with the encouragement to study the relevant materials for the topics with low performance.

## 4. Evaluation methodology

As mentioned earlier, the focus of this work is to evaluate the self-assessment tool in the domain of Microeconomics for high-school students. The main research questions are:

Research question 1: Is the tool usable?
Research question 2: Is the tool suitable for the field of school Microeconomics?
Research question 3: Does the tool improve student engagement?
Research question 4: Does the tool help students to improve their knowledge?

To answer these questions, we designed the following concrete tests:
1. Evaluation of the usability of the tool with the aid of teachers.
2. Evaluation of the suitability of the tool for learning and especially in the domain of school Microeconomics with the aid of teachers.
3. Implementation and evaluation of self-assessment tests with the aid of students.

All these steps have been implemented with the aid of teachers of economics and students who take Microeconomics classes. The focus of the evaluation is on school microeconomics. The law of demand, price elasticity of demand, factors affecting demand, and income elasticity of demand have been taught to students who attend the last class of a Greek Senior High School. Students are 17 years old and attend a specialized class of Microeconomics compulsory for the University admission examinations. Hence, these topics are quite important for them as they will have to take year-end admission exams on them. Figure 3 shows the stages of the evaluation. The dimensions of usability and educational value of the tool are assessed in this model.

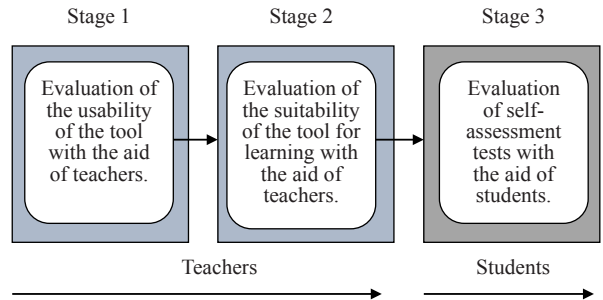

**Figure 3.** The stages of the tool evaluation

## 5. Evaluation experiments
### 5.1 *Usability evaluation & suitability for learning*

During these phases of the evaluation, answers to the first two research questions are sought. Usability is one of the vital requirements in software quality terms that a mobile learning system must provide (ISO, 1998). According to the ISO 9241-11 standard (ISO, 1998), usability is the capability of a product to be used by certain users to achieve specific objectives with effectiveness, efficiency, and satisfaction within a specific use context. The usability assessment of our tool can allow us to know the most important characteristics that are qualified when using the tool. For evaluating the usability of the tool and measuring its acceptability from school teachers, we employed the System Usability Scale (SUS) [https://www.usability.gov/how-to-and-tools/methods/system-usability-scale.html] method. With this questionnaire, we can have a reliable estimation of the usability of the system. The following questions, as defined in the SUS model, have been posed to 10 teachers (4 males and 6 females), with more than 10-year teaching experience who possessed at least moderate ICT skills:

1. I think that I would like to use this system frequently.
2. I found the system unnecessarily complex.
3. I thought the system was easy to use.
4. I think that I would need the support of a technical person to be able to use this system.
5. I found the various functions in this system were well integrated.
6. I thought there was too much inconsistency in this system.
7. I would imagine that most people would learn to use this system very quickly.
8. I found the system very cumbersome to use.
9. I felt very confident using the system.
10. I needed to learn a lot of things before I could get going with this system.

Teachers had to use the system for developing a short three-question test with three different types of questions. Then, they had to take a test developed by another teacher. Hence, each teacher had the opportunity to experiment with the tool both as a test creator and as a test taker. At the beginning of the experiment, we held a brief ten-minute introduction to the application and its purposes. Then, the teachers were free to first develop a short test and then to run the quiz developed by another participant.

As the SUS method dictates, each question is answered on a 5-likert scored from 1 (Strongly disagree) to 5 (Strongly agree). The final score of this task is 83.25 which is a strong indication of the high usability of the system.

After the computation of the usability score, we held a short interview with each teacher to record their opinions and experiences and to identify potential shortcomings. First, we were interested in the suitability of the tool for formative assessments. All the teachers were quite positive about the appropriateness of the application and they additionally felt it was an important learning tool. They argued that it could help the students realize their knowledge or misconceptions per concept and they could get them prepared for the final assessment. They considered the application as user-friendly and close to the reality of the students. All the evaluators agreed that they could easily promote it on the students, as they are used in using mobile applications. The tests would be readily available when needed which would suit different learning styles and studying habits (e.g., just a quick check before a test or using it after studying a textbook, etc.).

The final part of the interview focused on the domains that such a tool would be fit for. The participants agreed that such a tool would be appropriate for all technical topics. As such, quizzes for Microeconomics with different kinds of questions, including graphs and animations could be implemented with the tool. One of the points raised though, is that the images and other multimedia material should be created by using other tools. Indeed, the question editor supports mainly the development of text questions and the integration of visual objects. However, this is a common practice in most content management systems (Lust et al., 2012). The teachers argued that the ability to connect questions with specific Microeconomics concepts (e.g., price elasticity of demand) helps students to realize their knowledge and specific misconceptions. Presenting the results per concept at the end of the test could really enlighten the students. Overall, they considered that the tool would be a quite useful service to their classes.

## 5.2 *Student evaluation*

The third phase of the evaluation focused on students and sought to answer the last two research questions. With the aid of the two most experienced teachers, who have participated in the previous phase, we formed several questions related to linear demand curves, price elasticity of demand and the income elasticity of demand. A pen and pencil quiz version was also developed. The test contained 30 questions. We avoided using questions with video content that could not be transformed for the pen and pencil version of the quiz. The quizzes were unchangeable, which means that students could not use any of the adapted options. The participants would have to attempt all the questions in both versions, should they wished to practice. As explained, these students have been taught various microeconomics topics in a class following a traditional lecture with a specific textbook.

From a pool of 83 16-year old students attending the third class at a senior Greek high school, we formed three classes of students with 10 students in each class. The first class of students achieved high marks, i.e. 18-20 out of 20 during the last semester on the topic of Microeconomics. The next class included students with marks from 15-17 and the third class 10 students with grades 12-14. The students were randomly selected by the initial 83-learner pool. Randomly again, 5 students of each class were selected and formed the 15 student 'app group' and the rest 15 students formed the 'pen-and-pencil group'. Each group was informed that they participate in an experiment with the aim to realize the studying habits of students in general. All the data collected would be anonymous and their actions would not affect their grades at any rates. Students had to use their assigned version of the quiz as they would like, if they wanted to. There was no restriction. They were all informed that in a week they would have to take a test in the class, and so they could practice. The 'app group' could see the answers once the test was completed. The second group could bring the test back to a teacher to get feedback on the answers and to get additional explanations on their potential errors.

The application was available to the 'app group' in a period of 7-day. During this period, students could use the quizzes. For this period, we collected and processed various statistics about the usage of the application.

A primary concern was the measurement of the engagement of the students with the quizzes. All the students of the 'app group' run the quiz during the first day. There were also 33 more re-runs of the quiz during the following days. For the second group, computing this statistic was trickier. Six students definitely completed and submitted the test to their teacher seeking help on their answers. Four of these students belonged to the first class of the students with the higher past performance. The other 2 belonged to the second class who had been graded with 15-18 during the last semester. Five more students (2 of the first class, 2 of the second class, and 1 of the last class) of the 'pen-and-pencil group' informed us that they had completed the quiz, when we asked them. As a result, 100% of the 'app group' took the test one or more times, while in the second group 6/15 (40%) have definitely completed the test and 5/15 (33.33%) told us that they used the test.

During the class test (summative assessment) which contained questions similar to the self-assessment test, the average score of the 'app group' was 16.73 while the average score of the 'pen-and-pencil group' was 13.86. The results are statistically different as a 2-tailed t-test showed (p-value is 0.046568, the result is significant at $p < 0.05$). The two upper classes consisted of the students with the higher marks, had almost equal averages. This was expected as they are more dedicated and motivated students, as their marks reveal at least. Significant differences were observed in the other two classes. For example, the lowest scores were 12/20 and 13/20 in the third class of the 'app group' and 5/20 and 9/20 in the second group. We asked the students who scored worse in the test, if they took advantage of the material they had prior to the test. The students of the 'app group' said that they ran the test 2-3 times and studied the textbook once, while the students of the 'pen-and-pencil group' had not completed the quiz and they also had a quick study of the textbook. Further, we asked all the students of the second group if they would prefer to use the application and think they would eventually use it. All but one of the students of the second group were more enthusiastic towards the use of the application and they declared that they would use it. The more skeptical student raised a concern about the provided feedback. This student had filled the pen and pencil quiz, (s)he submitted it to the teacher and had a counseling session with the teacher to get explanations on various questions (s)he had. That was an excellent observation that helps us to advance our tool by supporting the addition of specific explanations per question option. Also, we could facilitate the communication of students and teachers by allowing the automatic forwarding of the results to a teacher and the ability to ask specific questions either by email or in a forum integrated into the tool. This way we could improve the shortcomings pointed out by the student. Teachers have not identified the lack of feedback as an issue, which may be due to the predominant belief that class tests are mainly summative. The tool supports both formative and summative assessments but in the present study the main aims are to examine the usability of the tool and the increase in their engagement. So, this feature will definitely be included in a future version of the tool.

The students of the 'app group' were asked whether they think that the application helped them to realize their true knowledge. All of them agreed or strongly agreed with this suggestion. They pointed out that the tool showed the erroneous questions at the end and the associated concepts, which led them to revisit specific sections of the textbook.

Finally, we distributed the SUS questionnaire to all the students of the 'app group' to have an estimation of the usability of the tool by the students as well. Again we had a score higher than 80 which is again a strong indication of the usability of the tool.

## 6. Discussion and conclusions

In the previous sections, we presented an application for self-assessments. We ran two different evaluation tests with the aid of teachers and students to answer specific research questions.

The first question concerned with the usability of the tool. Applying a standard method for realizing the easiness of use of the tool, we can positively argue that no serious usability issues have arisen. Users found the tool easy to use, with no prerequisites, and without the need for constant support by an expert.

The second research question was posed directly to the teachers and indirectly to the students. All the teachers agreed that the tool is suitable for many domains, especially the technical ones, including the field of school Microeconomics. In other words, all the domains which can be measured with quantifiable quizzes may be supported through our tool. Students indirectly supported this argument through their universal engagement with the tool. Everyone who had access to the tool used it once or more time to run the implemented quiz. 14/15 of the students who had not used the tool would prefer to have access to the application. Overall, there are strong indications of the suitability of the tool for school Microeconomics.

The answers to the last two research questions are again supportive of our work. All the students used the tool once or more times, and those with no access wished to have used it. Adolescents and even young children are almost addicted to their mobile phones. So it is probably easier to approach learning through these devices. Intelligent and usable applications would help in this direction. Students who used the application had better performance than the others, primarily because even the less dedicated students 'played' with the application and eventually studied or understood some of the testing material. Helping the less involved students to the learning process is a major advantage and is of crucial importance to most learning tools and serious games. Our tool seems to help in this direction.

This paper presented a self-assessment tool and considered its suitability for learning Microeconomics concepts. However, it is noteworthy that this study evaluated only some adaptation features of this tool as it is difficult to test all the features of an even moderately complex learning application. This is going to be researched in this school year through a series of evaluation tests aiming at understanding the influence of the adaptation to the specific aims of the students. For instance, instead of re-taking the entire test, one could be more interested in questions of increased difficulty or which concern specific concepts. This could eventually improve the engagement especially if question items for various disciplines are created and thus the utilization of the tool would be increased.

To sum up, there are strong indications of the acceptability of the application by both students and teachers.

The evaluation experiments with a satisfactory number of participants strengthen this belief. The engagement of the students and the improvement in their understanding are also supportive. Some intelligent observations were made by the participants which will be considered during the next development phase. The basic advantage of our work is that it improves the motivation and participation of students in the learning process.